\documentclass[preprint2]{aastex62}
\usepackage{amssymb}
\usepackage{amsmath}
\usepackage{gensymb}
\usepackage{url}
\usepackage{epsfig}
\graphicspath{{./}{figures/}}

\shorttitle{Scattering of energetic electrons by whistlers}
\shortauthors{Roberg-Clark et al.}

\begin{document}

\title{Scattering of energetic electrons by heat-flux-driven whistlers in flares}

\author[0000-0001-5280-2644]{G.~T.~Roberg-Clark}
\thanks{Work was carried out at the \\ University of Maryland, College Park}
\affil{Max Planck Institute for Plasma Physics, EURATOM Association, Wendelsteinstr. 1, 17491 Greifswald, Germany}
\email{gareth.roberg-clark@ipp.mpg.de}

\author[0000-0002-9150-1841]{O.~Agapitov}
\affil{Space Sciences Laboratory, University of California Berkeley, Berkeley, CA, USA}
\email{oleksiy.agapitov@gmail.com}

\author[0000-0002-9150-1841]{J.~F.~Drake}
\affil{Department of Physics, University of Maryland, College Park, MD 20740, USA}
\affil{Institute for Research in Electronics and Applied Physics, University of Maryland, College Park, MD 20742, USA}
\affil{Institute for Physical Science and Technology, University of Maryland, College Park, MD 20742, USA}
\affil{Joint Space-Science Institute (JSI), College Park, MD 20742, USA}
\email{drake@umd.edu}

\author[0000-0002-5435-3544]{M.~Swisdak}
\affil{Department of Physics, University of Maryland, College Park, MD 20740, USA}
\affil{Institute for Research in Electronics and Applied Physics, University of Maryland, College Park, MD 20742, USA}
\affil{Joint Space-Science Institute (JSI), College Park, MD 20742, USA}
\email{swisdak@umd.edu}

\begin{abstract}
The scattering of electrons by heat-flux-driven whistler waves is
explored with a particle-in-cell (PIC) simulation relevant to the
transport of energetic electrons in flares. The simulation is
initiated with a large heat flux that is produced using a kappa
distribution of electrons with positive velocity and a cold return
current beam. This system represents energetic electrons escaping from
a reconnection-driven energy release site. This heat flux system
drives large amplitude oblique whistler waves propagating both along
and against the heat flux, as well as electron acoustic waves. While the waves are dominantly driven by the low energy electrons, including the cold return current beam, the energetic electrons resonate with and are scattered by the whistlers on time scales of the order of a hundred electron cyclotron
times. Peak whistler amplitudes of $\tilde{B} / B_{0} \sim 0.125$ and angles of $\sim 60 \degree$ with respect to the background magnetic field are observed. Electron perpendicular energy is increased while the field-aligned electron heat flux is suppressed. The resulting scattering mean-free-paths of energetic electrons are small compared with the typical scale size of energy release sites in flares, which might lead to the effective confinement of energetic electrons
that is required for the production of very energetic particles.

\end{abstract}


\section{Introduction} \label{sec:intro}

Magnetic reconnection is the driver of explosive energy release in the
sun's corona \citep{Forbes1988} and in plasma environments throughout
the universe \citep{Michel1994}. In solar flares electrons can have energies exceeding a MeV \citep{Lin2003,Gary2018} and in astrophysical environments can reach a PeV \citep{Abdo2011}. An important question that arises in building a model for electron acceleration in the sun and elsewhere is how electrons remain within the region where
magnetic energy is being released for a sufficiently long time to reach the energies seen in observations.

The magnetic energy release rate in flares is controlled by
magnetic reconnection in which the upper limit on the rate of
reconnection is of order of $0.1 \: V_A$ where $V_A$ is the local Alfv\'en
speed. For typical parameters ($B\sim 50 \: \text{G}$, $n\sim 10^{9} \: \text{cm}^{-3}$) $V_{A}$ is around $3000$ km/s. The scale size of magnetic energy release in large solar flares can reach $10^{4}$ km so energy release rates are tens of seconds. In the absence of a confinement mechanism the transit time of
a relativistic electron out of this region is of the order of
$0.03 \: \text{s}$. Consistent with these estimates, the decay time of hard X-ray
emission from flares exceeds the transit time of energetic electrons
across the source by two orders of magnitude
\citep{Masuda1994,Krucker2007,Krucker2010}.

Thus, some mechanism for electron confinement is necessary to hold
electrons in the energy release zone for a long enough time to reach
relativistic energies. Because the Larmor radius of even relativistic
electrons is small ($\simeq 7 \: \text{cm}$ for the given numbers), one possibility is a form of magnetic confinement. However, magnetic field lines that link flare energy release sites end either on the chromosphere or the solar wind so
confinement is not effective unless the electrons can mirror. Significant mirroring of energetic electrons would require that their velocity have a large component perpendicular to the ambient magnetic field. Recent models of electron acceleration during reconnection suggest that electron energy gain is mostly parallel to
$\mathbf{B}$ and is dominated either by parallel electric fields or Fermi
reflection \citep{Drake2006,Dahlin2014,Dahlin2016}. The consequence is that electron
distribution functions are strongly anisotropic even in 3D
reconnecting systems, which become turbulent with the development of
multiple x-lines and chaotic magnetic fields \citep{Dahlin2017}.

There is evidence from RHESSI spacecraft observations that the
electron energy flux in the source regions of flares can exceed that
measured at the chromosphere by up to an order of magnitude
\citep{Simoes2013}. It has been suggested that double layers driven by
the return current of cold electrons interacting with ambient ions
would cause reflection of some hot electrons. However, since the
resulting potential drop across the double layers scales like
$T_{eh} \ll m_ec^2$ \citep{Li2013,Li2014} with $T_{eh}$ the temperature
of hot escaping electrons, the amplitude of double layers is not
sufficient to trap relativistic electrons. Further, the direct
measurement of radio emission from gyro-synchrotron emission from
energetic electrons in flares requires significant energy in the
perpendicular motion of electrons \citep{Gary2018}. Thus, if the
dominant acceleration mechanism of electrons in flares is parallel to
the ambient magnetic field, a mechanism is required to scatter the
parallel energy into perpendicular energy.

A potential scattering mechanism for energetic electrons is via oblique
whistler waves (e.g. \cite{Artemyev2012,Artemyev2014}). The whistler resonance condition with electrons is given by $\omega -k_\parallel v_\parallel -n\Omega_{e}/\gamma=0$, where $\Omega_{e}=eB_{0}/(m_{e}c)$ is the electron cyclotron frequency, $n$ is an integer that can take on positive and negative values
\citep{Krall1986} and $\gamma$ is the relativistic Lorentz
factor. For typical waves with $kd_e\sim 1$ the resonant velocities are
given by $v_\parallel\sim nV_{Ae}$, where $d_e$ is the electron skin
depth and $V_{Ae}$ is the electron Alfv\'en speed. Scattering by an oblique whistler at each resonance can be efficient. Furthermore when overlap of resonances occurs, scattering can be strongly increased for a large fraction of electrons. \citep{Roberg-Clark2016,Karimabadi1992a}. Kinetic simulations with boundary
conditions that impose a fixed temperature jump along an ambient
magnetic field have established that in high $\beta$ systems the
fluctuating magnetic field from whistlers is comparable to the initial
ambient magnetic field, which is sufficient to strongly scatter
electrons. This limits their effective streaming velocity to the
whistler phase speed, which in a $\beta \sim 1$ system is of order
$V_{Ae}$.

In these previous works treating high-$\beta$ systems \citep{Roberg-Clark2018a,Roberg-Clark2018b,Komarov2018a} the imposed electron distribution functions were Maxwellians with specified
temperature jumps that drove a heat flux. Here we consider a system in
which the heat flux arises from electrons with a $\kappa$ distribution
propagating in one direction and a cold electron beam that produces a
return current. In this system oblique whistlers develop that
propagate both along and against the heat flux. Oblique whistlers propagating along the direction of the heat flux are driven at early time by the ``fan'' instability of the (anomalous) $n=-1$ cyclotron resonance \citep{Kadomtsev1968,Haber1978,Fulop2006,Krafft2010,Vasko2019,Verscharen2019}. At later time the growth of parallel-propagating electron acoustic waves facilitates the extraction of energy from the bulk of the hot electrons in tandem with the oblique whistlers. Whistlers propagating against the heat flux are driven by the Landau resonance with the return current electrons. Both classes of whistlers resonate with and scatter the most energetic electrons in the tail of the $\kappa$
distribution, reducing their heat flux and substantially increasing their velocity perpendicular to the magnetic field.

\section{Simulation Method}
We carry out a two-dimensional (2D) simulation using the PIC code $\tt{p3d}$ \citep{Zeiler2002} to model energetic electrons and a cold return current electron beam. $\tt{p3d}$ calculates particle trajectories using the relativistic Newton-Lorentz equations and the electromagnetic fields are advanced using Maxwell's equations. An initially uniform magnetic field $\mathbf{B_{0}}=B_{0} \mathbf{\hat{x}}$ threads the plasma. $v_{x}$ is therefore the parallel velocity and $v_{y}$ and $v_{z}$ are the perpendicular velocities, while $y$ is the perpendicular spatial coordinate. Boundary conditions are periodic in $x$ and $y$. The initial particle distribution function has two components. The first is a parallel ($v_{x} >0$) bi-$\kappa$ distribution with temperatures $T_{h,x} \gg T_{h,\perp}$,
\begin{equation}
\begin{split}
    f_{h,\kappa} = \frac{ n_{0} \Gamma(\kappa + 1) } {\pi^{3/2} \theta^{2}_{h,\perp} \theta_{h,x} \kappa^{3/2} \Gamma(\kappa - 1/2)} \times \\ \left[ 1 + \frac{v^{2}_{x}}{\kappa \theta^{2}_{x}} + \frac{v^{2}_{\perp}}{\kappa \theta^{2}_{\perp}} \right]^{-(\kappa+1)} \Theta(v_{x}).
\end{split}
\label{eqn:fhot}
\end{equation}
$n_{0}$ is the initial density of each of the electron components, $\Gamma$ is the gamma function,

\begin{equation} 
\theta_{hx,\perp} = \left[(\kappa-3/2)/\kappa \right]^{1/2}V_{T{hx,\perp}}
\end{equation}
are the effective thermal speeds, $V_{Thx,\perp}=\sqrt{2T_{hx,\perp}/m_{e}}$ are the regular thermal speeds, $\Theta(v_{x})$ is the Heaviside step function and $\kappa$ is a parameter that tunes the steepness of the nonthermal tail of the distribution.

The second electron component is the cold return current beam (moving against $B_{0}$) which takes the form of a drifting isotropic Maxwellian,
\begin{equation}
\begin{split}
f_{c} = \frac{n_{0}}{\pi^{3/2}} \frac{e^{-[(v_{x} + v_{d})^2 + v_{\perp}^2]/v_{Tc}^2}}{v_{Tc}^3(1+\text{erf}(v_{d}/v_{Tc}))}\Theta(-v_{x}),
\end{split}
\label{eqn:fcold}
\end{equation}
where $V_{Tc}=\sqrt{2T_{c}/m_{e}}$ is the cold thermal speed and $v_{d}$ is a drift speed that ensures zero net current ($\langle v_{\parallel} \rangle =0$) in the initial state while the error function $\text{erf}(v_{d}/v_{Tc})$ makes the density of hot and cold particles equal. This choice is motivated by observations of flares \citep{Krucker2010,Oka2013} suggesting that the population of energetic electrons is large.

For the simulation presented we chose $\kappa=4$, $T_{x}=20 \: T_{\perp}=20 \: T_{c}$, $\beta_{e0h}=8\pi n T/B^{2}_{0} \sim 8\pi n_{0}T_{hx}/B_{0}^{2} = 2$. While this is a relatively high $\beta$ for the corona , a $\beta \sim 1$ system is consistent with specific flare observations \citep{Krucker2010}, is inferred more generally from rough equipartition of energy release between heated and energetic ions and electrons in flares \citep{Emslie2005,Emslie2012}, and is seen in reconnection simulations \citep{Dahlin2017}. The large value of $T_{hx}/T_{h\perp}$ is motivated by 3D PIC simulations of reconnection in \cite{Dahlin2017} showing $P_{\parallel}/P_{\perp} \sim 100$. This system is marginally stable with respect to the fluid firehose criterion, i.e. $\beta_{\parallel} - \beta_{\perp} \lesssim 2$.

The simulation domain lengths are $L_{x}=L_{0}=163.84\ d_{e}$ and $L_{y}=L_{0}/2$, where $d_{e} = c/\omega_{pe}$ is the electron skin depth and $\omega_{pe}=(4\pi n_{0}e^{2}/m_{e})^{1/2}$ is the electron plasma frequency. Other parameters in the simulation include $\omega_{pe}/\Omega_{e0} = 5 \sqrt{2}$, and $T_{hx}/(m_{e}c^{2})=0.02$, which sets $v_{Thx}/c = 1/5$. The characteristic velocity of whistlers depends on the wavelength but has an upper limit that scales with the electron Alfv\'en speed  $V_{A,e}=d_{e}\Omega_{e0}$, where $\Omega_{e0}=eB_{0}/m_{e}c$. Ions, with mass ratio $m_{i}/m_{e}=1600$, are initialized with a Maxwellian distribution of temperature $T_{i0}=T_{eh}/2$ and do not play a significant role in the simulation. The simulation uses $560$ particles per species per cell, has a grid of $4096$ by $2048$ cells, and is run to the time $t\Omega_{e0} = 1332$.

\section{Simulation Results}

\begin{figure}
\plotone{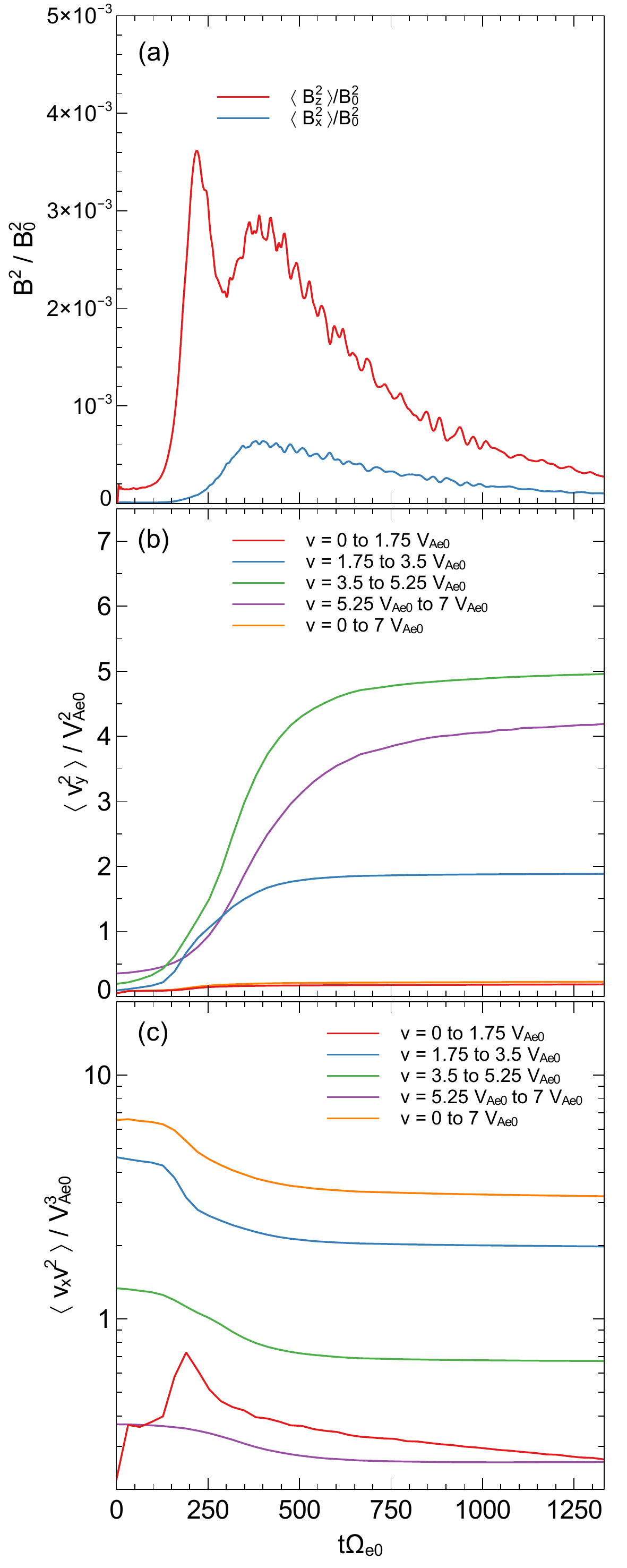}
\caption{(a) Fluctuation amplitudes $\langle B^{2}_{z}  \rangle$ and $\langle \delta B^{2}_{x} \rangle$ as a function of time. (b) Scattering of electrons at different energies. Four of the curves are plots of $\langle v^{2}_{y} \rangle$ in evenly spaced ranges with velocity width $v=1.75 V_{Ae0}$, where $v=\sqrt{v^{2}_{x}+v^{2}_{y}}$. Each curve is normalized by the number of particles in their velocity range. The fifth curve is the same quantity but for the entire velocity range and nearly overlaps that of the lowest energy bin. (c) Same as (b) but for the energy flux $\langle v_{x}v^{2} \rangle$ without normalization by particle number and plotted on a log scale.}
\label{fig:timeplots}
\end{figure}

The initial distribution function [the total of Eqs.~(\ref{eqn:fhot}) and (\ref{eqn:fcold})] drives magnetic fluctuations unstable in the system. Shown in Fig.~\ref{fig:timeplots}(a) is the time evolution of the box-averaged magnetic fluctuation energies $\langle B^{2}_{z} \rangle$ and $\langle \delta B^{2}_{x} \rangle$, where $\delta B_{x} = B_{x} - B_{0}$. The energy in $B_{z}$ peaks at around $t\Omega_{e0}=190$ after a fast growth phase while a second peak occurs at $t\Omega_{e0} \sim 375$. The magnetic fluctuations damp significantly by the end of the simulation, approaching early-time noise levels.

\begin{figure}
\plotone{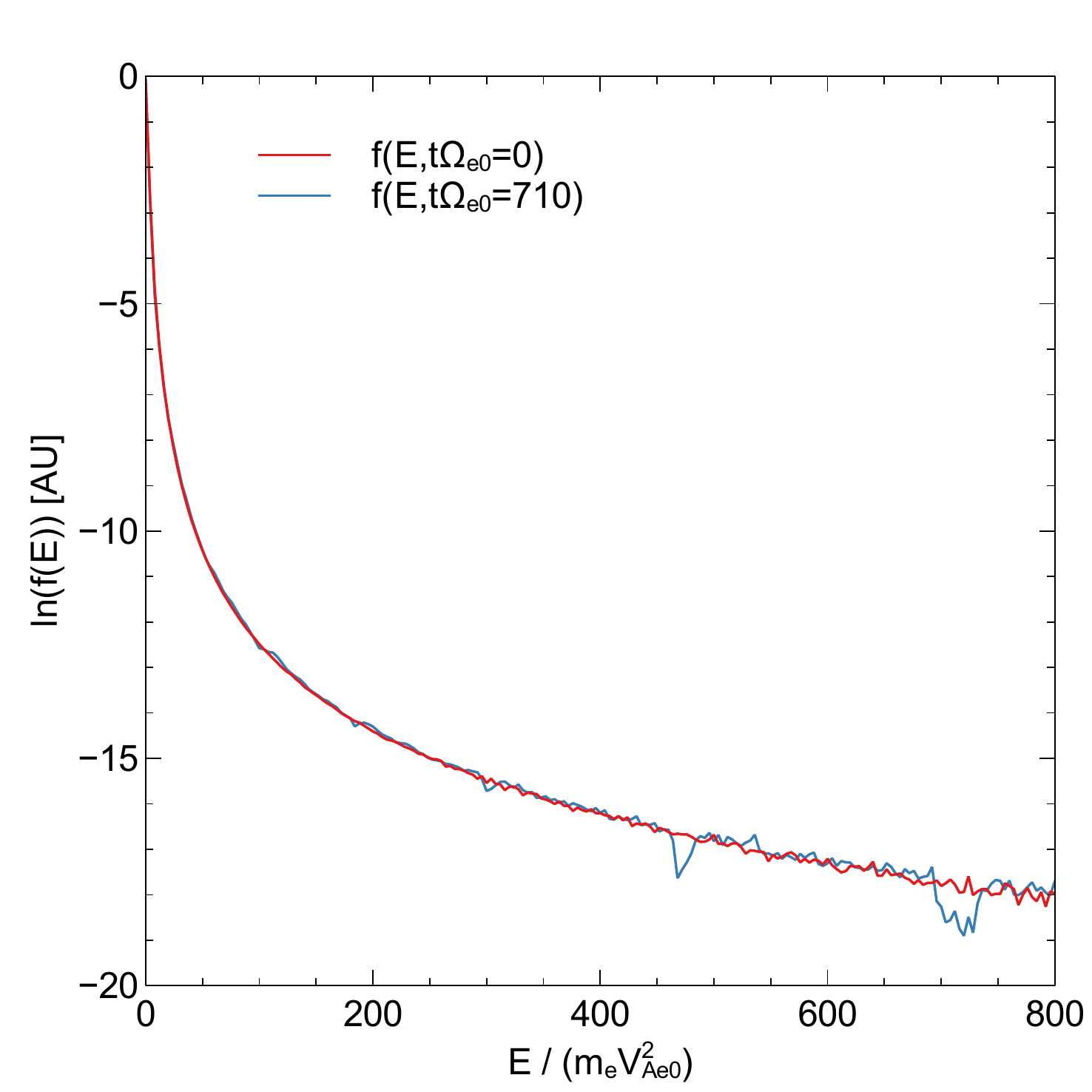}
\caption{Electron energy distributions $\ln[f(E)]$ at $t\Omega_{e0}=0$ and $710$. $E=(\gamma-1)m_{e}c^{2}$ is the relativistic kinetic energy.}
\label{fig:energydist}
\end{figure}

To establish that the growth of the fluctuations is associated with the scattering of electrons, we track the time dependence of the electron distribution function $f_{e}(v_{x},v_{y})$, with $v_{z}$ as well as the two spatial coordinates averaged out. We use $v_{x}$ and $v_{y}$ as proxies for the parallel and perpendicular velocities since $\mathbf{B} \sim B_{0}\hat{x}$ and the distribution remains gyrotropic $(v_{y} \sim v_{z} \sim v_{\perp})$ for the duration of the simulation (not shown). We divide velocity space into ranges of width $\delta v=1.75 V_{Ae0}$, treating $v=\sqrt{v^{2}_{x}+v^{2}_{y}}$ as the total velocity. In Fig.~\ref{fig:timeplots}(b) the time dependence of  $\langle v^{2}_{y} \rangle$ is shown for each velocity range with labels indicating the velocity intervals. For each of the energy ranges $\langle v^{2}_{y} \rangle$ increases with time. Figure \ref{fig:energydist} reveals that the distribution of energy $E=(\gamma - 1)m_{e}c^{2}$ is almost unchanged during the course of the simulation since the initial and late-time distributions (at $t\Omega_{e0}=0$ and $710$) basically overlap for most energies (and only vary slightly at high energy where there are few particles). Thus, the increase in $\langle v^{2}_{y} \rangle$ in time corresponds to a decrease in $\langle v^{2}_{x} \rangle$ so the dominant scattering is in pitch angle. The sharpest increase in $\langle v^{2}_{y} \rangle$ occurs at around $t\Omega_{e0}=300$. The characteristic time over which $\langle v^{2}_{y} \rangle$ increases is roughly $100 \: \Omega^{-1}_{e0}$, which we take to be the scattering time of the electrons. Saturation of $\langle v^{2}_{y} \rangle$ (the ``rollover") takes place for a long period of time starting around $t\Omega_{e0}=400$ and continues until the end of the simulation.

Scattering of the energy from $v_{x}$ to $v_{y}$ also reduces the energy flux, which is shown in Fig.~\ref{fig:timeplots}(c) where we plot the time dependence of $\langle v_{x}v^{2} \rangle$ for the same energy bins as in Fig. \ref{fig:timeplots}b. Note that the data is presented on a log scale and we do not divide by the number of particles in each velocity range for this quantity. Curves representing most of the velocity ranges show a monotonic decrease of $\langle v_{x}v^{2} \rangle$ with time, in some cases resulting in a drop by a factor of $2$. An exception is the curve corresponding to $0 < v/V_{ae0} < 1.75$ which increases until it peaks at $t\Omega_{e0} \sim 190$ and then drops off.

\begin{figure*}
\plotone{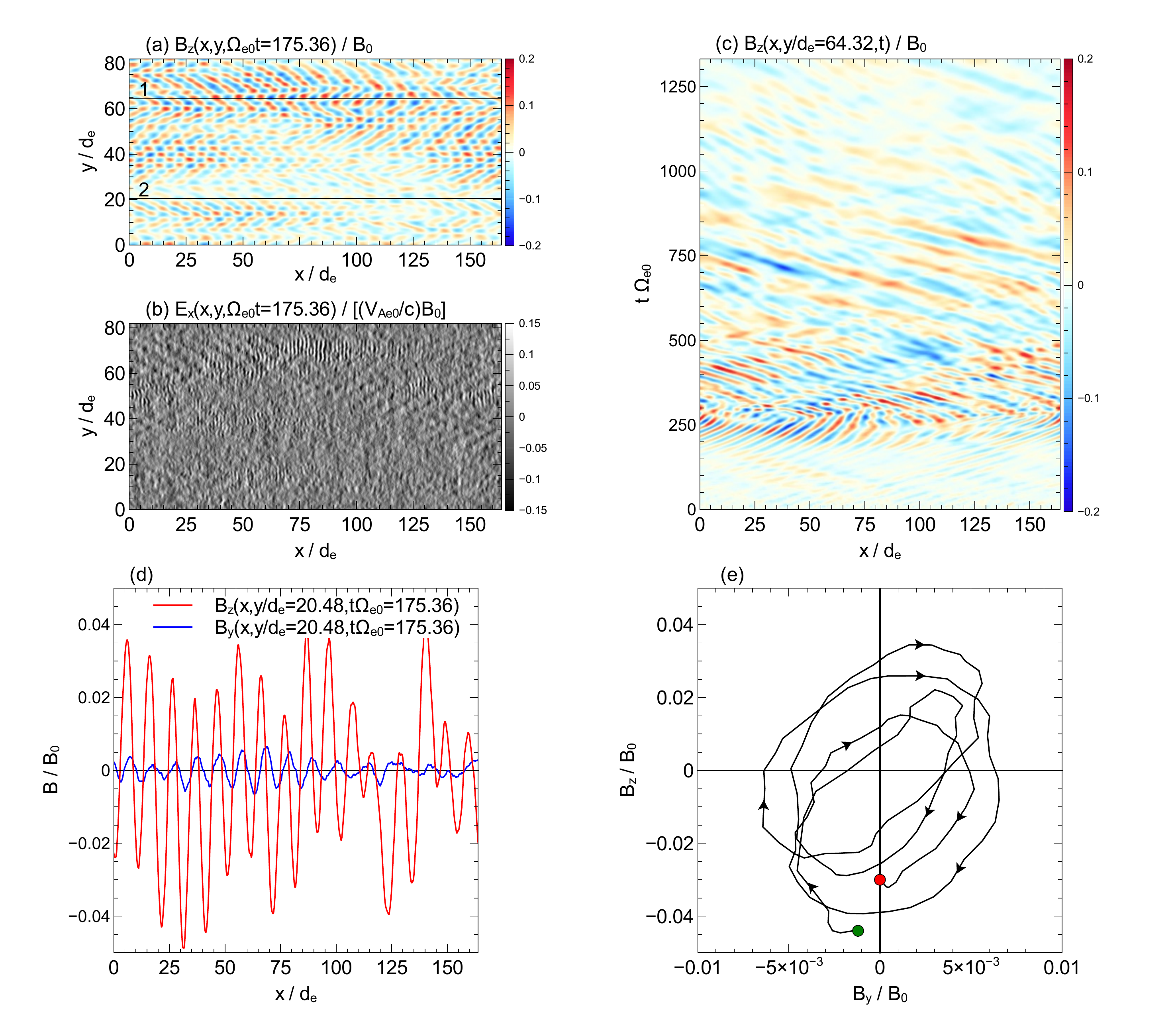}
\caption{Magnetic and electric fields in the simulation. (a) 2D image of $B_{z}$ at $t\Omega_{e0}=175$. (b) The same as (a) but for $E_{x}$. (c) Spacetime diagram in $x-t$ space for the entire simulation at the location $y/d_{e}=64.32$ (line ``1'' in (a)). (d) Line plots of the perpendicular magnetic fields $B_{y}$ and $B_{z}$ at $y/d_{e}=20.48$ (line ``2'' in (a)). (e) Hodogram ($B_{y}$ vs. $B_{z}$) at line ``2'' but from $x/d_{e}=40.96$ to $81.92$.}
\label{fig:fields}
\end{figure*}

To establish the nature of the fluctuations that develop in the simulation we show in Fig. \ref{fig:fields} the structure and motion of the out-of-plane magnetic fluctuations $B_{z}$ and the parallel electric field fluctuations $E_{x}$. Figure \ref{fig:fields}a is a 2D plot of $B_{z}$ at early time, $t\Omega_{e0}=175$, during the first growth phase. The fluctuations travel at angles of roughly $60$ degrees relative to $B_{0}$ as predicted in \cite{Verscharen2019} for whistler scattering of energetic particles in a low-$\beta$ system. The Verscharen et al. result is constrained by the location of the $n=-1$ resonance and a minimization of Landau damping, which is ensured if the whistler phase speed is large enough compared to the core electron thermal speed. In our simulation, however, the particle distribution is flat in $v_{\parallel}$ near the whistler phase speed and so the whistler does not suffer from large Landau damping in the $\beta \sim 1$ case. Rather, the angle of $60 \degree$ maximizes the wave-particle interaction for the $n=-1$ resonance (equation \ref{eqn:trappingwidths}). The crossed interference pattern that emerge in the simulation seem to be a consequence of the symmetry between the plus and minus $y$ directions. For every oblique wave generated with a velocity in the plus $y$ direction, there will be a companion wave with a velocity in the negative $y$ direction, which leads to a standing-wave-like pattern. Spatial inhomogeneities in the wave pattern are probably due to variations in noise levels in the system at early time. 

A spacetime diagram of $B_{z}$ at a cut at $y=64.32 \: d_{e}$ is shown in Fig. \ref{fig:fields}c. Waves moving in the $+\hat{x}$ direction become visible around $t\Omega_{e0} = 200$ (recall the first peak in the fluctuating $B_{z}$ in Fig.~\ref{fig:timeplots}b at $\Omega_{e0}t \sim 175$) but are beginning to slow down and reverse their direction by this time. The parallel phase speed of these waves at early time ($t\Omega_{e0} \sim 100$) is $v_{p,x}/ V_{Ae0}=(\omega/k_{x})/V_{Ae0} \sim +0.5$ while at later time the leftward-propagating waves move at roughly $v_{p,x}/V_{Ae0} \sim -0.5$ with noticeably longer parallel wavelengths.

In Fig. \ref{fig:fields}d are plots of $B_{y}$ and $B_{z}$ along a cut at $y/d_{e}=20.48$ revealing what appears to be a $90$ degree phase shift between the two components. We also show a hodogram of these quantities from $x/d_{e} = 40.96$ to $81.92$ in Fig. \ref{fig:fields}e demonstrating right-handed elliptical polarization, confirming that these are indeed whistlers.

The cold plasma dispersion relation for whistlers (neglecting the displacement current) is 
\begin{equation}
    \omega=\frac{|k_{x}|kd_{e}^{2}\Omega_{e}}{1+k^{2}d_{e}^{2}}
    \label{eqn:disp}
\end{equation}
with $k_{x} > 0$ ($<0$) corresponding to a ``rightward'' (``leftward'') whistler propagating along (against) $B_{0}$. From Eq.~(\ref{eqn:disp}) the phase speed of $0.5 \: V_{Ae0}$ seen at early time in the simulation corresponds to a wave with $kd_{e}\simeq 1$, a value that is consistent with the wavelength of fluctuations in Fig.~\ref{fig:fields}a. The fluctuations in the electric field at $t\Omega_{e0}=175$ (Fig.~\ref{fig:fields}b) show crossed patterns similar to those in \ref{fig:fields}a. These are the parallel electric fields of the rightward moving oblique whistlers, which are significant for large oblique angles and for $kd_{e} \gtrsim 1$. In Fig.~\ref{fig:fields}b there are shorter-wavelength, parallel-propagating modes with $k \lambda_{De} \sim 10$ and $\omega \simeq 0.4 \: \omega_{pe}$ which are electron acoustic waves (EAWs) [$\lambda_{De}=V_{Te}/(\sqrt{2}\omega_{pe})$ is the electron Debye length]. The waves grow through the Landau resonance between the thermal speeds of the hot and cold electron components and reduce the relative drift between those components \citep{Gary1985,Agapitov2018,Vasko2018}. These modes are of large amplitude, $\tilde{E}/((V_{Ae0}/c)B_{0}) \sim 0.3$ (note Fig. \ref{fig:fields}b has its color scale capped at $0.15$ to bring out the whistler signal).

\begin{figure}
\plotone{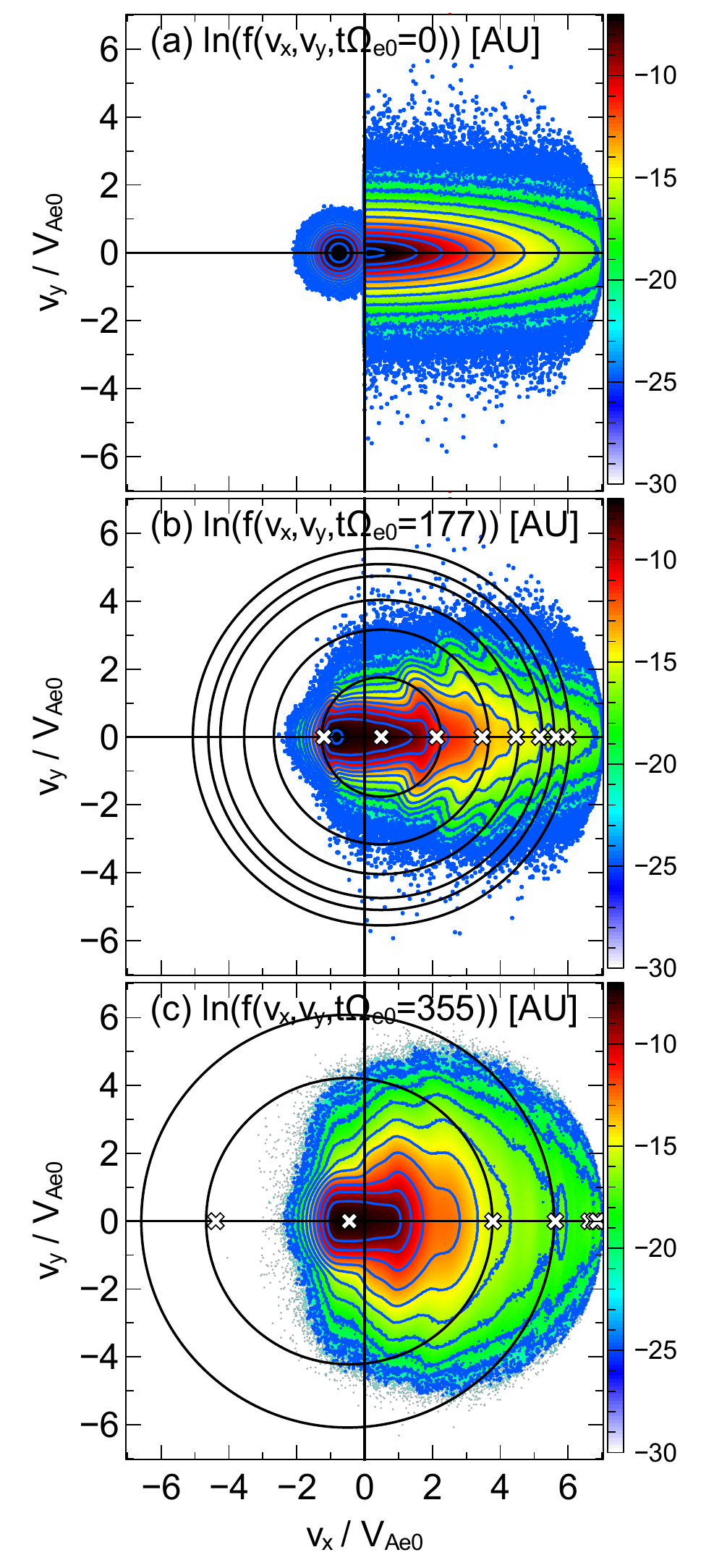}
\caption{Electron distribution functions in the $v_{x}-v_{y}$ phase space at different times, shown both in color and with contours. (a) The imposed distribution at $t=0$ with a return current Maxwellian for $v_{x}<0$ and energetic, anisotropic bi-$\kappa$ for $v_{x}>0$. (b) At $t\Omega_{e0}=177$ horn-like figures have emerged as a result of scattering. Intersections of the resonant surfaces $n=1,0,-1,-2,-3,-4,-5$ with the $v_{x}=0$ axis (white crosses, equation \ref{eqn:rel-res}) are shown with the constant-energy surfaces $\gamma_{0}=1.035,1.125,1.25,1.45,1.6,$ and $1.8$ (solid black lines, equation \ref{eqn:rel-energy}). (c) At $t\Omega_{e0}=355$, the $n=-1$ through $n=5$ resonances for the leftward wave are shown along with the energy surfaces $\gamma_{0}=1.3$ and $1.9$.}
\label{fig:vxvydist}
\end{figure}

Figures \ref{fig:vxvydist}a-c illustrate the scattering of the electron distribution function as time proceeds in the simulation. The color plot shows the contours of constant $f(v_{x},v_{y})$. At $t=0$ (Fig.~\ref{fig:vxvydist}a) the large discontinuity in $f$ at $t=0$ separates the return current beam with $v_{x} < 0$ and the $\kappa$ distribution with $v_{x} > 0$. At $t\Omega_{e0}=177$ (Fig.~\ref{fig:vxvydist}b) the distribution develops horn-like structures near $v_{x} / V_{Ae0} \simeq -1, 1.8, 3,$ and $3.5$ that demonstrate that particles from the initial distribution near the $v_{y}$=0 axis have been scattered to higher $v_{y}$ and lower $v_{x}$. The largest number of scattered particles is in the structure at $v_{x}/V_{Ae0}=1.8$. The discontinuity in Fig. \ref{fig:vxvydist}a has been filled in and the contours of the distribution are fairly flat in the vicinity of the whistler phase speed $v_{p,x}/V_{Ae0}\sim 0.5$. We attribute the flattening to large-amplitude electrostatic fluctuations that quickly grow up and damp in the simulation (not shown). When $t\Omega_{e0}=355$ (\ref{fig:vxvydist}c), the distribution is significantly more isotropic in the $v_{x}>0$ half-plane. While some particles have been scattered to $v_{x}<0$, most of the scattering seems to be limited to $v_{x}>0$, suggesting that $\langle v_{y}^{2} \rangle$ saturates in Fig.~\ref{fig:timeplots}a because the $v_{x}>0$ half-plane has become nearly isotropic. Figure \ref{fig:vxvydist}c is therefore representative of the late-time structure of the distribution function.

\section{Resonances}
\label{sec:resonances}

\begin{figure*}
\plotone{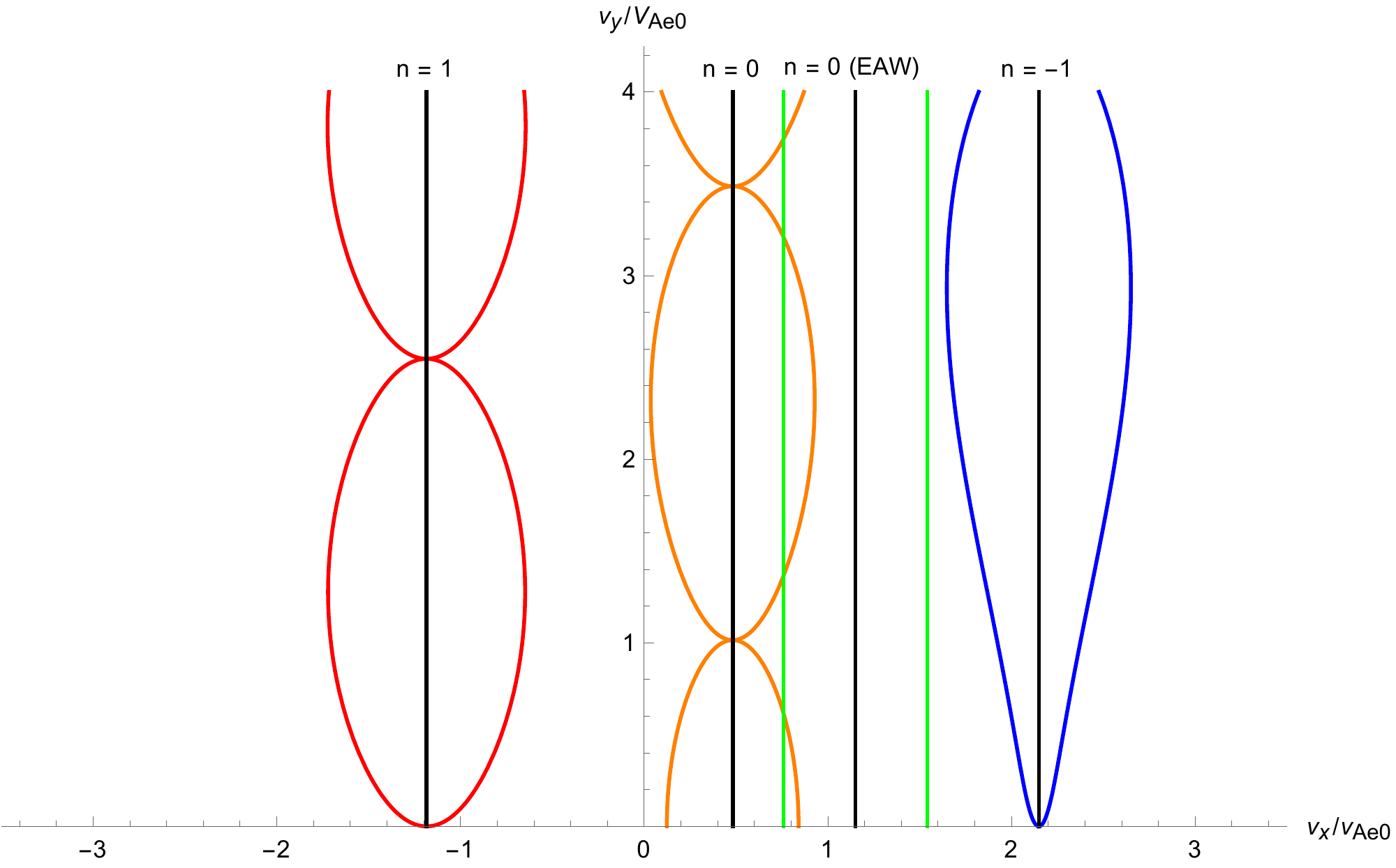}
\caption{Electron trapping widths. The $n=0$ (orange), $n=1$ (red) and $n=-1$ (blue)  resonances for the rightward-propagating whistler at peak amplitude ($\tilde B/B_{0}=0.125$,$k_{x}d_{e}=0.6,k_{y}d_{e}=1$), calculated using expression (\ref{eqn:trappingwidths}). The EAW $n=0$ trapping width (green) is overlaid for $k_{x}d_{e}=3.92$, $\tilde{E}=0.3$, and $v_{p,x}/V_{Ae0} = 1.17$.}
\label{fig:trapping}
\end{figure*}

To explain scattering in the simulation we invoke the basic theory of resonant interaction between oblique whistlers and electrons (see e.g. \cite{Roberg-Clark2016} and references therein). We write the resonance condition as 
\begin{equation}
    \omega - k_{x}v_{x,r} - \frac{n\Omega_{e0}}{\gamma} = 0
    \label{eqn:res}
\end{equation}
where $n=0,\pm1,\pm2,...$, $v_{x,r}$ is the parallel resonant velocity, and $\gamma=(1-v_{r}^{2}/c^{2})^{-1/2}$ with $v_r$ the total electron velocity. We first discuss the non-relativistic case ($\gamma=1$) for which the resonance is represented by a vertical line $v_{x}=v_{x,r}$ in the $v_{x}-v_{y}$ space. In the long-wavelength limit, $k^{2}d^{2}_{e} \ll 1$, the electric field of a single whistler is eliminated in a frame moving along $B_{0}$ at the speed $(v_{p,x}=\omega/k_{x})$. Energy conservation, $(v_{x}-v_{p,x})^{2} + v_{y}^{2} = \text{const}$, then requires that particle orbits lie on circles centered around $v_{x}=v_{p,x}$ as they oscillate in the fields of the whistler. When $kd_{e} \gtrsim 1$ the whistler retains a finite electric field in its frame and particle energy is not exactly conserved \citep{Karimabadi1990,Karimabadi1992a}. The nonlinear trapping width associated with a resonance $n$ can be calculated from the electron equation of motion in the whistler frame. Using the linearized cold plasma dispersion relation to obtain the whistler field components (neglecting the displacement current) we find the parallel trapping width of the $n$th resonance to be
\begin{widetext}
\begin{equation}
\Delta v_{x,n}= 2\sqrt{2} \left| \frac{\Omega_{e}}{k_{x}} \frac{\tilde{B}}{B_{0}} \left[ \frac{\omega}{|k_{x}|} \frac{k_{y}}{k} k^{2}d_{e}^2 J_{n}
+ \frac{v_{\perp 0}}{2} \left( \left( \frac{k}{|k_{x}|}-1 \right) J_{n+1}
+ \left( \frac{k}{|k_{x}|} + 1 \right) J_{n-1} \right) \right] \right|^{1/2}
\label{eqn:trappingwidths}
\end{equation}
\end{widetext}
where $J_{n}(k_{y}v_{y0} / \Omega_{e0})$ is the Bessel function of order n, $v_{y0}$ is the initial perpendicular velocity of the particle as it becomes resonant, and the whistler eigenvector amplitude $\tilde{B}$ is that of $\tilde{B_{y}}$. Figure \ref{fig:trapping} shows a schematic of the trapping widths for the $n=0,\pm 1$ resonances with $k_{x}d_{e}=0.6$, $k_{y}d_{e}=1$, which is the location of the peak in the spectrum obtained from the FFT of $B_{z}$ at $t\Omega_{e0}=177$ (not shown). The peak whistler amplitude of $\tilde{B}/B_{0}=0.125$ is chosen to evaluate the trapping widths. The trapping widths in $v_{x}$ are comparable to $V_{Ae0}$ for $v_{y 0}/V_{Ae0} \gtrsim 1.5$.

Figure \ref{fig:trapping} establishes that the $n=-1$ resonance strongly scatters particles with $v_{x}>0$, producing the large horn-like feature near $v_{x}/V_{Ae0} \lesssim 2$ in Fig.~\ref{fig:vxvydist}b and transferring energy from the particles to the wave. The horn-like feature could also explain why the heat flux of the lowest-energy particles in Fig.~\ref{fig:timeplots} peaks for $t\Omega_{e0}\sim 100$ since initially higher-energy particles could be scattered into the lowest velocity bin. The $n=1$ resonance scatters some of the particles in the return current beam with $v_{x}/V_{Ae0} \sim  -1$ to higher $v_{y}$ and smaller $v_{x}$ (note the small horn-like feature at this location in Fig.~\ref{fig:vxvydist}b). This interaction has a damping effect on the wave but the drive from $n=-1$ still dominates since the number of upscattered cold return current electrons is small. Initially the distribution function is flat near the whistler phase speed (Fig. \ref{fig:vxvydist}) so any effect on the energy of the whistler by the $n=0$ resonance is small. As a result the fan instability associated with the $n=-1$ resonance is what drives the wave at early time.

The rightward-moving electron acoustic waves (EAWs) in Fig. \ref{fig:fields}b are parallel-propagating, so only the Landau resonance is significant. It produces a trapping width 
\begin{equation}
    \Delta v_{x,0}= 2\sqrt{2} \sqrt{\frac{\Omega_{e}}{k_{x}} \frac{\tilde{E}}{B_{0}}c}
    \label{eqn:EAW-trap}
\end{equation}
where $\tilde{E}$ is the EAW amplitude. Setting $k_{x}d_{e}=3.92$ and $\tilde{E}=0.3$ (the peak amplitude observed) we find that the EAW Landau resonance width (indicated by green vertical lines) overlaps with that of the whistler (orange curves) for a region of the phase space near $v_{y}/V_{Ae0}=2.4$ and nearly overlaps that of the $n=-1$ resonance (blue curves) near $v_{y}/V_{Ae0} \lesssim 3$. Such overlap should lead to irreversible diffusion in phase space, dragging particles towards $v_{x}=0$ as seen over time in Fig.~\ref{fig:vxvydist}(a-c). If the overlap is not quite reached between the two whistler resonances and that of the EAW, chaotic orbits will still set in as overlap is approached. A small spread in $k$ can also shift the centroids of the traps and bridge the gap between the resonances. Importantly, a particle's energy can be reduced by a whistler and an EAW acting in tandem. The $n=-1$ hornlike feature in Fig. \ref{fig:vxvydist}b has been shifted to the EAW phase speed $v_{p,x}/V_{Ae0}=1.17$ in Fig. \ref{fig:vxvydist}c, pulling some particles into the whistler $n=0$ resonance at late time. This is a simple, graphical way to describe the nonlinear interaction of whistlers and EAWs \citep{Agapitov2018,Vasko2018,Drake2015}. The interaction between the whistler $n=-1$ resonance and the EAW resonance is analogous to the upscattering of high-energy particles and subsequent driving of plasma waves via a bump-on-tail instability as discussed in \cite{Haber1978}.

As particles lose energy and diffuse towards $v_{x}=0$ they also lose momentum. To maintain zero net current the system generates an inductive field that slows the average drift speed of the return current beam. The leftward moving whistler grows by drawing energy from the return current beam via the Landau resonance at the phase speed $v_{p,x}/V_{Ae0} \sim -0.5$. The flattening of the distribution across $v_{x}$ in Fig.~\ref{fig:vxvydist}c is evidence that this is taking place. This leftward whistler does not grow at early time because of damping of this wave by particles at the $n=1$ resonance \citep{Verscharen2019}. At late time the electron distribution at this resonance has been flattened by the action of the rightward propagating whistler so growth of the negative propagating wave through the Landau resonance takes place. Note that the large number of particles in the return current means the dominant energy transfer to the leftward wave is through the Landau resonance, so it is a beam-driven mode. 

\section{Relativistic effects}
Since there are very few particles at high energy for the initial distribution function chosen for the simulation, the high-energy component basically act as test particles. Their feedback on the wave dynamics is small. For these particles relativistic effects are important. In the relativistic case ($\gamma \gg 1$) the resonant surfaces and constant energy surfaces become elliptical in the $v_{x}-v_{\perp}$ plane \citep{Omidi1982,Karimabadi1990}. Trapping occurs around the intersection point of these two ellipses \citep{Karimabadi1990}, assuming that to lowest order in wave amplitude the particle energy is conserved in the wave frame.

Intersection of the resonance ellipse with the $v_{x}$ axis is given by
\begin{equation}
    v_{r,x} = \frac{v_{p,x}}{1+\alpha_{n}^{2}} \pm \sqrt{\frac{\alpha^{2}_{n}}{1+\alpha^{2}_{n}} \left(c^{2} - \frac{v_{p,x}^{2}}{1+\alpha^{2}_{n}} \right)}
    \label{eqn:rel-res}
\end{equation}
where $\alpha_{n}=n\Omega_{e0}/(k_{x}c)$ and $v_{p,x}$ is calculated from Eq.~(\ref{eqn:disp}). The constant-energy ellipses \citep{Karimabadi1990} are given by
\begin{equation}
     \frac{ \left( v_{x} - v_{x,c} \right)^{2} }{R/\left( \gamma_{0}^{2} + v_{p,x}^{2}/c^{2} \right)} +  \frac{v_{y}^{2}}{R/\gamma_{0}^{2}} = 1 
    \label{eqn:rel-energy}
\end{equation}
with $R=c^{2}(\gamma_{0}^{2}-1) + v_{p,x}^{2}/(\gamma_{0}^{2} + v_{p,x}^{2}/c^{2})$, $ v_{x,c} = v_{p,x}/(\gamma_{0}^{2} + v_{p,x}^{2}/c^{2})$ and $\gamma_{0}$ the initial Lorentz factor of a particle in the lab frame. For the whistlers in the simulation with $v_{p,x} \ll c$ the $v_{x}$ and $v_{y}$ axes of the ellipses are nearly equal. In the highly relativistic limit ($\gamma_{0} \rightarrow \infty$) the surface is a circle centered around $v=0$ with radius $c$ as expected. 

Figure \ref{fig:vxvydist}b displays the location of the resonance intersections for $n=1$ through $n=-5$ using (\ref{eqn:rel-res}) and nearby energy surfaces using (\ref{eqn:rel-energy}) assuming $k_{x}d_{e}=0.6$ for the rightward moving wave. The energy surfaces, the horn-like structures, and the resonance intersections line up surprisingly well, justifying the use of the relativistic theory and further confirming that the waves are whistlers. The $n=-1$ through $n=5$ resonances are also shown for the leftward wave ($k_{x}d_{e}=-0.2$, $k_{y}d_{e}=0.6$) in Fig.~\ref{fig:vxvydist}c, along with two energy surfaces. Since the leftward wave is long-wavelength, only the Landau resonance acts on low-energy particles. Scattering by the normal resonances $n=1,2$ likely aids in the diffusion of high-energy particles although the bulk of the scattering has already taken place by the time the leftward wave has a large amplitude. The most significant relativistic effect is the location of the resonances in Fig.~\ref{fig:vxvydist}. Expressions for the relativistic trapping widths, which we do not invoke here, can be found in \cite{Karimabadi1990}.

\section{Discussion}

We have found that energetic electrons escaping from a flare-like system with $\beta \sim 1$, which is expected from magnetic reconnection \citep{Dahlin2017}, efficiently drive whistler waves that scatter the escaping electrons. The whistlers pitch-angle scatter high-energy electrons on a rapid time scale of hundreds of cyclotron periods by means of cyclotron resonances, suppressing energy flux and increasing the perpendicular velocities of electrons. This is a local mechanism which can operate under the generic conditions of a reconnection-driven flare. Moderate reduction (up to a factor of two) of the field-aligned electron energy flux occurs. Since this scattering tends to increase perpendicular velocity, electrons will more effectively mirror when they encounter small-scale magnetic fluctuations in the corona or if they are accelerated towards the sun where the ambient magnetic field is stronger. Thus, this scattering mechanism will facilitate the confinement of energetic electrons in energy release sites in flares which is required for electrons to reach the relativistic velocities seen in observations \citep{Lin2003,Krucker2010,Gary2018}.

Although our model was designed to study flares, we point out some possible implications for transport into the outer corona and ultimately the solar wind. The fluctuations in our simulation damp out after scattering is complete. Nevertheless, the heat flux associated with the initial electron distribution is permanently reduced. We suggest on the basis of the present simulations that oblique waves could have grown to large amplitude either in the corona or in the solar wind in the outer reaches of the corona, scattered the electrons to reduce the heat flux and then died away, leaving a remnant reduced heat flux. We thus propose that the electron energy flux produced in the corona as a result of reconnection or other mechanisms could be suppressed by oblique whistlers, leading to marginally stable electron distributions that then propagate outward, leaving no trace of the self-generated turbulence that limited the heat flux. It has been shown that distribution functions can stream ballistically over several mean free paths before collisions dominate and the distribution becomes Maxwellian-like \citep{Malkov2017}. Since the mean free path of the solar wind is roughly $1 \text{AU}$, the streaming of heat-flux-carrying distributions out of the corona is thus feasible.

There is mounting evidence that the electron heat flux in the solar wind is limited by whistlers in regions with large plasma $\beta$ \citep{Tong2018}. However, the large-amplitude, oblique whistlers that would limit the heat flux have not been measured in the solar wind at $1$ AU \citep{Wilson2013,Lacombe2014,Stansby2016,Tong2019a,Tong2019,Kuzichev2019}, although some observations have implied effective pitch-angle scattering by whistlers with small angles relative to $B_{0}$ \citep{Kajdic2016}.  Oblique whistler scattering of the solar wind strahl into the halo is currently under active investigation (e.g. \cite{Verscharen2019}, \cite{Boldyrev2019}, \cite{Vasko2019}).

Our simulation addresses the $\beta \sim 1$ limit of flares. However, for a lower $\beta$ simulation, whistler growth would be suppressed since less free energy would be available \citep{Roberg-Clark2018b}. Scattering would still be expected to occur and the mode would retain a finite angle to $B_{0}$. Since the electron Alfv\'en speed would increase relative to the thermal speed ($\beta=(V_{T}/V_{Ae})^{1/2}$) the whistler phase speed might lie farther out in the tail of the distribution, leading to scattering of the highest energy particles.

We caution that our simulation suffers from the usual constraints of particle-in-cell simulations such as small simulated domains and short time scales compared to relevant scales in the corona. The model heat flux distribution function we use also contains a very sharp gradient near $v_{x}=0$. However, sharp gradients are likely to form during flares, during which particles are rapidly accelerated and our intent is to explore the possible mechanisms at play that would limit electron escape. Our model is relevant to $\beta \sim 1$ weakly collisional plasmas and could also be applied to transport in astrophysical coronae and low-luminosity accretion flows.

\acknowledgments

The authors acknowledge support from NSF Grant No. PHY1805829, from NASA
grant NNX17AG27G and from the FIELDS team of the
Parker Solar Probe (NASA Contract No. NNN06AA01C). GTRC was funded by an Ann G. Wylie Dissertation Fellowship from the University of Maryland, College Park. OA and JFD were supported by NASA grant 80NNSC19K0848. OA was partially supported by NSF grant number 1914670. JFD acknowledges partial support from NSF Grant No. PHY1748958 at the Kavli Institute for Theoretical Physics at UCSB. This research used resources of the National Energy Research Scientific Computing Center, a DOE Office of Science User Facility supported by the Office of Science of the U.S. Department of Energy under Contract No.DE-AC02-05CH11231. Simulation data is available upon request.

\bibliography{library}

\begin{thebibliography}{}
\expandafter\ifx\csname natexlab\endcsname\relax\def\natexlab#1{#1}\fi
\providecommand{\url}[1]{\href{#1}{#1}}

\bibitem[{Abdo {et~al.}(2011)Abdo, Ackermann, Ajello, Allafort, Baldini,
  Ballet, Barbiellini, Bastieri, Bechtol, Bellazzini, Berenji, Blandford,
  Bloom, Bonamente, Borgland, Bouvier, Brandt, Bregeon, Brez, Brigida, Bruel,
  Buehler, Buson, Caliandro, Cameron, Cannon, Caraveo, Casandjian, {\c{C}}elik,
  Charles, Chekhtman, Cheung, Chiang, Ciprini, Claus, Cohen-Tanugi, Costamante,
  Cutini, D'Ammando, Dermer, de~Angelis, de~Luca, de~Palma, Digel, {do Couto e
  Silva}, Drell, Drlica-Wagner, Dubois, Dumora, Favuzzi, Fegan, Ferrara, Focke,
  Fortin, Frailis, Fukazawa, Funk, Fusco, Gargano, Gasparrini, Gehrels,
  Germani, Giglietto, Giordano, Giroletti, Glanzman, Godfrey, Grenier, Grondin,
  Grove, Guiriec, Hadasch, Hanabata, Harding, Hayashi, Hayashida, Hays, Horan,
  Itoh, J{\'{o}}hannesson, Johnson, Johnson, Khangulyan, Kamae, Katagiri,
  Kataoka, Kerr, Kn{\"{o}}dlseder, Kuss, Lande, Latronico, Lee,
  Lemoine-Goumard, Longo, Loparco, Lubrano, Madejski, Makeev, Marelli,
  Mazziotta, McEnery, Michelson, Mitthumsiri, Mizuno, Moiseev, Monte, Monzani,
  Morselli, Moskalenko, Murgia, Nakamori, Naumann-Godo, Nolan, Norris, Nuss,
  Ohsugi, Okumura, Omodei, Ormes, Ozaki, Paneque, Parent, Pelassa, Pepe,
  Pesce-Rollins, Pierbattista, Piron, Porter, Rain{\`{o}}, Rando, Ray, Razzano,
  Reimer, Reimer, Reposeur, Ritz, Romani, Sadrozinski, Sanchez, Parkinson,
  Scargle, Schalk, Sgr{\`{o}}, Siskind, Smith, Spandre, Spinelli, Strickman,
  Suson, Takahashi, Takahashi, Tanaka, Thayer, Thompson, Tibaldo, Torres,
  Tosti, Tramacere, Troja, Uchiyama, Vandenbroucke, Vasileiou, Vianello,
  Vitale, Wang, Wood, Yang, \& Ziegler}]{Abdo2011}
Abdo, A.~A., Ackermann, M., Ajello, M., {et~al.} 2011, Science, 331, 739 LP

\bibitem[{Agapitov {et~al.}(2018)Agapitov, Drake, Vasko, Mozer, Artemyev,
  Krasnoselskikh, Angelopoulos, Wygant, \& Reeves}]{Agapitov2018}
Agapitov, O., Drake, J.~F., Vasko, I., {et~al.} 2018, Geophysical Research
  Letters, 45, 2168

\bibitem[{Artemyev {et~al.}(2012)Artemyev, Agapitov, Breuillard,
  Krasnoselskikh, \& Rolland}]{Artemyev2012}
Artemyev, A., Agapitov, O., Breuillard, H., Krasnoselskikh, V., \& Rolland, G.
  2012, Geophysical Research Letters, 39, 2007

\bibitem[{Artemyev {et~al.}(2014)Artemyev, Vasiliev, Mourenas, Agapitov,
  Krasnoselskikh, Boscher, \& Rolland}]{Artemyev2014}
Artemyev, A.~V., Vasiliev, A.~A., Mourenas, D., {et~al.} 2014, Geophysical
  Research Letters, 41, 5727

\bibitem[{Boldyrev \& Horaites(2019)}]{Boldyrev2019}
Boldyrev, S., \& Horaites, K. 2019, arXiv:1908.01902

\bibitem[{Dahlin {et~al.}(2014)Dahlin, Drake, \& Swisdak}]{Dahlin2014}
Dahlin, J.~T., Drake, J.~F., \& Swisdak, M. 2014, Physics of Plasmas, 21,
  092304

\bibitem[{Dahlin {et~al.}(2016)Dahlin, Drake, \& Swisdak}]{Dahlin2016}
---. 2016, Physics of Plasmas, 23

\bibitem[{Dahlin {et~al.}(2017)Dahlin, Drake, \& Swisdak}]{Dahlin2017}
---. 2017, Physics of Plasmas, 24, 092110

\bibitem[{Drake {et~al.}(2015)Drake, Agapitov, \& Mozer}]{Drake2015}
Drake, J.~F., Agapitov, O.~V., \& Mozer, F.~S. 2015, Geophysical Research
  Letters, 42, 2563

\bibitem[{Drake {et~al.}(2006)Drake, Swisdak, Che, \& Shay}]{Drake2006}
Drake, J.~F., Swisdak, M., Che, H., \& Shay, M.~A. 2006, Nature, 443, 553

\bibitem[{Emslie {et~al.}(2005)Emslie, Dennis, Holman, \& Hudson}]{Emslie2005}
Emslie, A.~G., Dennis, B.~R., Holman, G.~D., \& Hudson, H.~S. 2005, Journal of
  Geophysical Research: Space Physics, 110, 1

\bibitem[{Emslie {et~al.}(2012)Emslie, Dennis, Shih, Chamberlin, Mewaldt,
  Moore, Share, Vourlidas, \& Welsch}]{Emslie2012}
Emslie, A.~G., Dennis, B.~R., Shih, A.~Y., {et~al.} 2012, Astrophysical
  Journal, 759, doi:10.1088/0004-637X/759/1/71

\bibitem[{Forbes(1988)}]{Forbes1988}
Forbes, T.~G. 1988, in Astrophysics and Space Science Library book series
  (ASSL, volume 143)

\bibitem[{F{\"{u}}l{\"{o}}p {et~al.}(2006)F{\"{u}}l{\"{o}}p, Pokol, Helander,
  \& Lisak}]{Fulop2006}
F{\"{u}}l{\"{o}}p, T., Pokol, G., Helander, P., \& Lisak, M. 2006, Physics of
  Plasmas, 13, 062506

\bibitem[{Gary {et~al.}(2018)Gary, Chen, Dennis, Fleishman, Hurford, Krucker,
  McTiernan, Nita, Shih, White, \& Yu}]{Gary2018}
Gary, D.~E., Chen, B., Dennis, B.~R., {et~al.} 2018, The Astrophysical Journal,
  863, 83

\bibitem[{Gary \& Tokar(1985)}]{Gary1985}
Gary, S.~P., \& Tokar, R.~L. 1985, Physics of Fluids, 28, 2439

\bibitem[{Haber {et~al.}(1978)Haber, Huba, Palmadesso, \&
  Papadopoulos}]{Haber1978}
Haber, I., Huba, J.~D., Palmadesso, P., \& Papadopoulos, K. 1978, The Physics
  of Fluids, 21, 1013

\bibitem[{Kadomtsev \& Pogutse(1968)}]{Kadomtsev1968}
Kadomtsev, B., \& Pogutse, O. 1968, Soviet Journal of Experimental and
  Theoretical Physics, 26, 1146

\bibitem[{Kajdi{\v{c}} {et~al.}(2016)Kajdi{\v{c}}, Alexandrova, Maksimovic,
  Lacombe, \& Fazakerley}]{Kajdic2016}
Kajdi{\v{c}}, P., Alexandrova, O., Maksimovic, M., Lacombe, C., \& Fazakerley,
  A.~N. 2016, The Astrophysical Journal, 833, 172

\bibitem[{Karimabadi {et~al.}(1990)Karimabadi, Akimoto, Omidi, \&
  Menyuk}]{Karimabadi1990}
Karimabadi, H., Akimoto, K., Omidi, N., \& Menyuk, C.~R. 1990, Physics of
  Fluids B, 2, 606

\bibitem[{Karimabadi {et~al.}(1992)Karimabadi, Krauss-Varban, \&
  Terasawa}]{Karimabadi1992a}
Karimabadi, H., Krauss-Varban, D., \& Terasawa, T. 1992, Journal of Geophysical
  Research, 97, 13853

\bibitem[{Komarov {et~al.}(2018)Komarov, Schekochihin, Churazov, \&
  Spitkovsky}]{Komarov2018a}
Komarov, S., Schekochihin, A.~A., Churazov, E., \& Spitkovsky, A. 2018, J.
  Plasma Phys, 84

\bibitem[{Krafft \& Volokitin(2010)}]{Krafft2010}
Krafft, C., \& Volokitin, A. 2010, Physics of Plasmas, 17, 102303

\bibitem[{Krall \& Trivelpiece(1986)}]{Krall1986}
Krall, N.~A., \& Trivelpiece, A.~W. 1986, {Principles of Plasma Physics} (San
  Francisco Press)

\bibitem[{Krucker {et~al.}(2010)Krucker, Hudson, Glesener, White, Masuda,
  Wuelser, \& Lin}]{Krucker2010}
Krucker, S., Hudson, H.~S., Glesener, L., {et~al.} 2010, The Astrophysical
  Journal, 714, 1108

\bibitem[{Krucker {et~al.}(2007)Krucker, White, \& Lin}]{Krucker2007}
Krucker, S., White, S.~M., \& Lin, R.~P. 2007, The Astrophysical Journal, 669,
  49

\bibitem[{Kuzichev {et~al.}(2019)Kuzichev, Vasko, Soto-Chavez, Tong, Artemyev,
  Bale, \& Spitkovsky}]{Kuzichev2019}
Kuzichev, I.~V., Vasko, I.~Y., Soto-Chavez, A.~R., {et~al.} 2019, The
  Astrophysical Journal, 882, 81

\bibitem[{Lacombe {et~al.}(2014)Lacombe, Alexandrova, Matteini, Santol{\'{i}}k,
  Cornilleau-Wehrlin, Mangeney, {De Conchy}, \& Maksimovic}]{Lacombe2014}
Lacombe, C., Alexandrova, O., Matteini, L., {et~al.} 2014, The Astrophysical
  Journal, 796

\bibitem[{Li {et~al.}(2013)Li, Drake, \& Swisdak}]{Li2013}
Li, T.~C., Drake, J.~F., \& Swisdak, M. 2013, The Astrophysical Journal, 778,
  144

\bibitem[{Li {et~al.}(2014)Li, Drake, \& Swisdak}]{Li2014}
---. 2014, The Astrophysical Journal, 793, 7

\bibitem[{Lin {et~al.}(2003)Lin, Krucker, Hurford, Smith, Hudson, Holman,
  Schwartz, Dennis, Share, Murphy, Emslie, Johns-Krull, \& Vilmer}]{Lin2003}
Lin, R.~P., Krucker, S., Hurford, G.~J., {et~al.} 2003, The Astrophysical
  Journal, 595, 69

\bibitem[{Malkov(2017)}]{Malkov2017}
Malkov, M.~A. 2017, Physical Review D, 95, 1

\bibitem[{Masuda {et~al.}(1994)Masuda, Kosugi, Hara, Tsuneta, \&
  Ogawara}]{Masuda1994}
Masuda, S., Kosugi, T., Hara, H., Tsuneta, S., \& Ogawara, Y. 1994, Nature,
  371, 495

\bibitem[{Michel(1994)}]{Michel1994}
Michel, F.~C. 1994, The Astrophysical Journal, 431, 397

\bibitem[{Oka {et~al.}(2013)Oka, Ishikawa, Saint-Hilaire, Krucker, \&
  Lin}]{Oka2013}
Oka, M., Ishikawa, S., Saint-Hilaire, P., Krucker, S., \& Lin, R.~P. 2013,
  Astrophysical Journal, 764, 4

\bibitem[{Omidi \& Gurnett(1982)}]{Omidi1982}
Omidi, N., \& Gurnett, D.~A. 1982, Journal of Geophysical Research, 87, 2377

\bibitem[{Roberg-Clark {et~al.}(2016)Roberg-Clark, Drake, Reynolds, \&
  Swisdak}]{Roberg-Clark2016}
Roberg-Clark, G.~T., Drake, J.~F., Reynolds, C.~S., \& Swisdak, M. 2016,
  Astrophys. J. Lett., 830, L9

\bibitem[{Roberg-Clark {et~al.}(2018{\natexlab{a}})Roberg-Clark, Drake,
  Reynolds, \& Swisdak}]{Roberg-Clark2018a}
---. 2018{\natexlab{a}}, Phys. Rev. Lett., 120, 035101

\bibitem[{Roberg-Clark {et~al.}(2018{\natexlab{b}})Roberg-Clark, Drake,
  Swisdak, \& Reynolds}]{Roberg-Clark2018b}
Roberg-Clark, G.~T., Drake, J.~F., Swisdak, M., \& Reynolds, C.~S.
  2018{\natexlab{b}}, The Astrophysical Journal, 867, 154

\bibitem[{Sim{\~{o}}es {et~al.}(2013)Sim{\~{o}}es, Fletcher, Hudson, \&
  Russell}]{Simoes2013}
Sim{\~{o}}es, P.~J., Fletcher, L., Hudson, H.~S., \& Russell, A.~J. 2013,
  Astrophysical Journal, 777, 1

\bibitem[{Stansby {et~al.}(2016)Stansby, Horbury, Chen, \&
  Matteini}]{Stansby2016}
Stansby, D., Horbury, T.~S., Chen, C. H.~K., \& Matteini, L. 2016, The
  Astrophysical Journal, 829, L16

\bibitem[{Tong {et~al.}(2018)Tong, Bale, Salem, \& Pulupa}]{Tong2018}
Tong, Y., Bale, S.~D., Salem, C., \& Pulupa, M. 2018, arXiv:1801.07694

\bibitem[{Tong {et~al.}(2019{\natexlab{a}})Tong, Vasko, Artemyev, Bale, \&
  Mozer}]{Tong2019a}
Tong, Y., Vasko, I.~Y., Artemyev, A.~V., Bale, S.~D., \& Mozer, F.~S.
  2019{\natexlab{a}}, The Astrophysical Journal, 878, 41

\bibitem[{Tong {et~al.}(2019{\natexlab{b}})Tong, Vasko, Pulupa, Mozer, Bale,
  Artemyev, \& Krasnoselskikh}]{Tong2019}
Tong, Y., Vasko, I.~Y., Pulupa, M., {et~al.} 2019{\natexlab{b}}, The
  Astrophysical Journal Letters, 870, L6

\bibitem[{Vasko {et~al.}(2018)Vasko, Agapitov, Mozer, Bonnell, Artemyev,
  Krasnoselskikh, \& Tong}]{Vasko2018}
Vasko, I.~Y., Agapitov, O.~V., Mozer, F.~S., {et~al.} 2018, Physical Review
  Letters, 120, 195101

\bibitem[{Vasko {et~al.}(2019)Vasko, Krasnoselskikh, Tong, Bale, Bonnell, \&
  Mozer}]{Vasko2019}
Vasko, I.~Y., Krasnoselskikh, V., Tong, Y., {et~al.} 2019, The Astrophysical
  Journal, 871, L29

\bibitem[{Verscharen {et~al.}(2019)Verscharen, Chandran, Jeong, Salem, Pulupa,
  \& Bale}]{Verscharen2019}
Verscharen, D., Chandran, B. D.~G., Jeong, S.-Y., {et~al.} 2019,
  arXiv:1906.02832

\bibitem[{Wilson {et~al.}(2013)Wilson, Koval, Szabo, Breneman, Cattell, Goetz,
  Kellogg, Kersten, Kasper, Maruca, \& Pulupa}]{Wilson2013}
Wilson, L.~B., Koval, A., Szabo, A., {et~al.} 2013, Journal of Geophysical
  Research: Space Physics, 118, 5

\bibitem[{Zeiler {et~al.}(2002)Zeiler, Biskamp, Drake, Rogers, Shay, \&
  Scholer}]{Zeiler2002}
Zeiler, A., Biskamp, D., Drake, J.~F., {et~al.} 2002, Journal of Geophysical
  Research, 107, 1230

\end{thebibliography}



\end{document}